\documentclass[preprint,nofootinbib,floats,superscriptaddress,tightenlines,amsfonts,amsmath,amssymb]{revtex4}
\usepackage[colorlinks=true,
            linkcolor=blue,
            urlcolor=blue,
            citecolor=green,          
            bookmarks=true,
            bookmarksnumbered=true,
            breaklinks=true,
            pdfpagemode=Fullscreen,
            pdfstartview=FitBH]{hyperref}
\usepackage{graphicx}
\usepackage{bm}
\usepackage{xcolor}
\usepackage{multirow}
\usepackage{slashed}
\allowdisplaybreaks
\usepackage{siunitx}
\usepackage{diagbox}
\usepackage{ulem}
\usepackage{tablefootnote}
\usepackage{booktabs}
\usepackage{amsmath}
\usepackage[capitalise]{cleveref}
\newcommand{\calO}{{\cal O}}

\begin{document}

\baselineskip=16.5pt \parskip=5pt

\hspace*{\fill}

\title{ Revisiting models that enhance $B^+\to K^+ \nu\bar\nu$ in light of the new Belle II measurement}

\author{Xiao-Gang He}
\email{hexg@phys.ntu.edu.tw}
\affiliation{\small Tsung-Dao Lee Institute (TDLI) \& School of Physics and Astronomy (SPA),
Shanghai Jiao Tong University (SJTU), Shanghai 200240, China}
\affiliation{\small Department of Physics, National Taiwan University, Taipei 10617}

\author{Xiao-Dong Ma}
\email{maxid@scnu.edu.cn}
\affiliation{\small Key Laboratory of Atomic and Subatomic Structure and Quantum Control (MOE), Guangdong Basic Research Center of Excellence for Structure and Fundamental Interactions of Matter, Institute of Quantum Matter, South China Normal University, Guangzhou 510006, China} 
\affiliation{\small Guangdong-Hong Kong Joint Laboratory of Quantum Matter, Guangdong Provincial Key Laboratory of Nuclear Science, Southern Nuclear Science Computing Center, South China Normal University, Guangzhou 510006, China}

\author{German Valencia}
\email{german.valencia@monash.edu}
\affiliation{\small School of Physics and Astronomy, Monash University, Wellington Road, Clayton, VIC-3800, Australia\bigskip}

\begin{abstract}

Belle II has recently reported the new measurement ${\cal B}(B^+\to K^+\nu\bar\nu)=(2.3\pm0.7)\times 10^{-5}$ \cite{Belle-II:2023esi}
which is two times larger than their previous result (although consistent within errors) and about $2.7\,\sigma$ higher than  the SM prediction.
We re-examine new physics scenarios we have discussed previously which can enhance this rate to determine
if they can accommodate the higher value reported in the new measurement.
We use consistency with existing bounds on $B\to K^*\nu\bar\nu$, $b\to s \ell^+\ell^-$,
$B\to D^{(*)}\ell\bar\nu$ and $B_s$ mixing to limit possible explanations for the excess.
For the case of LFV neutrino couplings, we find that only two leptoquarks remain viable
requiring a large $C_{9^\prime}^{\tau\tau}=-C_{10^\prime}^{\tau\tau}$.
For models with different types of light dark matter particle pairs (scalar, fermion, or vector),
the preliminary $q^2$ distribution from Belle II, which shows that the excess appears mostly for bins with $3\leq q^2\leq7$ GeV$^2$ \cite{Belle-II:2023esi}, 
implies only the vector current operators with scalar or vector  dark matter particles with masses in the hundreds of MeV can match the anomaly. 

\end{abstract}

\maketitle

{\small\hypersetup{linkcolor=black}\tableofcontents}

\newpage

\section{Introduction}

The $B\to K^{(*)} \nu \bar \nu$ decays are amongst the cleanest modes to search for new physics
due to their well-controlled theoretical uncertainty in the standard model (SM).
A recent measurement by the Belle II collaboration finds a branching ratio \cite{Belle-II:2023esi} 
\begin{align}
{\cal B}(B^+ \to K^+ \nu \bar \nu)_{\tt exp}=(2.3 \pm 0.7) \times 10^{- 5},
\label{eq:newm}
\end{align}
which is higher than SM expectation by about $2.7\,\sigma$.
This result is also about twice as large, but consistent within errors,
with a previous Belle II combination 
${\cal B}(B^+\to K^+\nu\bar\nu)_{\tt exp}^{2021}= (1.1\pm0.4)\times 10^{-5}$ \cite{Dattola:2021cmw},
and with the new average 
${\cal B}(B^+\to K^+\nu\bar\nu)_{\tt exp}^{\tt ave}= (1.3\pm0.4)\times 10^{-5}$ \cite{Belle-II:2023esi}.
Currently, these numbers suggest consistency with the SM prediction (subtracting the so-called tree-level contribution)\cite{Becirevic:2023aov},
\begin{align}
{\cal B}(B^+\to K^+\nu\bar\nu)_{\tt SM} = (4.43\pm 0.31)\times 10^{-6},
\end{align}
but the new measurement is sufficiently intriguing to entertain the possibility of new physics
affecting this decay \cite{Asadi:2023ucx,Bause:2023mfe,Allwicher:2023syp,Athron:2023hmz,Felkl:2023ayn,Abdughani:2023dlr,Dreiner:2023cms}. 
Note that this number agrees with the one we use for our numerical study below,
obtained from {\tt flavio}, ${\cal B}(B^+\to K^+\nu\bar\nu)_{\tt SM}=(4.4\pm0.6)\times 10^{-6}$.
It has been noted that the CKM dependence of the SM prediction, which can hide new physics (NP),
can be bypassed by considering certain ratios instead \cite{Buras:2022qip}.\footnote{The SM value we use is a bit lower
than the one quoted in \cite{Belle-II:2023esi} due to the use of different CKM parameters: 
we consistently use  $|V_{tb}V^*_{ts}|=(3.93\pm 0.10)\times 10^{-2}$ whereas \cite{Parrott:2022zte},
quoted in  \cite{Belle-II:2023esi} uses  $|V_{tb}V^*_{ts}|=0.04185 (93)$.}

We have  previously studied three different new physics scenarios that can enhance the rate for this mode:
lepton flavour violating neutrino couplings (possibly induced by leptoquarks) \cite{He:2021yoz};
a light sterile neutrino \cite{He:2021yoz}; or other invisible light particles (dark matter) \cite{He:2022ljo}.
These three cases exploit the fact that the neutrinos (or their flavour) are not detected so they could be mimicked by other unseen particles. 

In this paper, we revisit those scenarios with the new (high) measurement in mind.
Specifically, we want to explore whether this large value of  ${\cal B}(B^+ \to K^+ \nu \bar \nu)$ (or indeed $B^+\to K^+ + \slashed{E}$)
can be accommodated in any of those scenarios while satisfying the existing upper bounds on ${\cal B}(B \to K^* \nu \bar \nu)$ (again, more generally, $B\to K^* + \slashed{E}$). All these NP scenarios introduce correlations with other modes that also need to be considered.

Specifically, we will base our study on the ratio,
\begin{align}
R^K_{\nu\nu} \equiv  
\frac{\mathcal{B}(B^+\to K^+\nu\bar\nu)}{\mathcal{B}(B^+\to K^+\nu\bar\nu)_{\tt SM}}, 
\end{align}
using the numerical values obtained from both the new Belle II measurement and from the new average reported in Fig.\,23 of \cite{Belle-II:2023esi}
\begin{subequations}
\label{eq:newrn}
\begin{align}
\left(R^K_{\nu\nu}\right)_{\tt new ~Belle~II}=& ~5.3\pm 1.7,
\\
\left(R^K_{\nu\nu}\right)_{\tt new ~average}=& ~3.0\pm 1.0.  
\end{align}
\end{subequations}
We will refer to the first number as the ``new $1\,\sigma$ range''. 
The rate predicted for the corresponding neutral and charged pseudoscalar meson modes is the same if isospin is conserved,
and the measurement of the neutral mode is at present less restrictive. To discuss correlations with $R^{\nu\nu}_{K^*}$ we will use two numbers:
\begin{align}
R^{\nu\nu}_{K^*}=&\frac{\mathcal{B}(B\to K^*\nu\bar\nu)}{\mathcal{B}(B\to K^*\nu\bar\nu)_{\tt SM}}  \leq 2.7 {\rm ~or~}  1.9.
  \label{eq:rkstar}
\end{align}
The first number (2.7) arises from the combined charged and neutral modes as directly quoted by the Belle collaboration \cite{Belle:2017oht}.
The second number reflects the fact that the predictions for the decay rates of the charged and neutral modes are the same in all the models we consider.
This suggests using the strongest experimental constraint, which in this case occurs for the neutral mode  \cite{Belle:2017oht}
\begin{align}
{\cal B}(B^0 \to K^{0*} \nu \bar \nu) \leq 1.8  \times 10^{- 5} {\rm~(90\%~C.L.)}.
\end{align}
In combination with the corresponding SM prediction, this results in the second number (1.9) in \cref{eq:rkstar}.

The overall picture suggested by these numbers is that the new Belle II measurement does not imply significant changes to the averages yet.
However, it invites us to entertain the possibility of a larger $R^{\nu\nu}_K$ and to study its implications for phenomenology,
in particular for the implied $R^{\nu\nu}_K>R^{\nu\nu}_{K^*}$. That is the subject of this paper.

\section{Altering the properties of the neutrinos}

We  parameterise possible new physics entering these decays through an effective Hamiltonian at the $b$ mass scale
with dimension six operators responsible for $b\to s \nu\bar \nu$ in the low energy effective field theory approach (LEFT) \cite{Jenkins:2017jig,Liao:2020zyx}.
The effective theory originates in extensions of the SM containing new particles at or above the electroweak scale
that have been integrated out but also allowing for the possibility of light right-handed neutrinos.
Specifically, we consider the effective Hamiltonian
\begin{align}
{\cal H}_{\tt NP} = -\frac{4G_F}{\sqrt{2}}V_{tb}V^\star_{ts}\frac{e^2}{16\pi^2}
&\sum_{ij}\left(C_{L}^{ij}{\cal O}_L^{ij}+C_{R}^{ij}{\cal O}_R^{ij}+
C_{L}^{\prime~ij}{\cal O}_L^{\prime~ij}+C_{R}^{\prime~ij}{\cal O}_R^{\prime~ij}\right ) +{\rm ~h.c.},
\label{eq:effHb}
\end{align}
including the operators 
\begin{subequations}
\label{eq:ops}
\begin{align}
{\cal O}_L^{ij}&= (\bar s_L\gamma_\mu b_L)(\bar\nu_i\gamma^\mu(1-\gamma_5)\nu_j),&
{\cal O}_R^{ij}&=(\bar s_R\gamma_\mu b_R)(\bar\nu_i\gamma^\mu(1-\gamma_5)\nu_j), 
\\
{\cal O}_L^{\prime~ij} &=(\bar s_L \gamma_\mu b_L )(\bar\nu_i\gamma^\mu(1+\gamma_5)\nu_j),&
{\cal O}_R^{\prime~ij} &=(\bar s_R\gamma_\mu b_R)(\bar\nu_i\gamma^\mu(1+\gamma_5)\nu_j)\;.
\end{align}
\end{subequations}
The Wilson coefficients in \cref{eq:ops} are defined so that they only contain NP contributions.
The SM contributes only to $C_L^{ii}$ and is accounted for separately \cite{Buchalla:1998ba,Brod:2010hi}
\begin{align}
C_{L\,\tt SM}=-\frac{X(x_t)}{s^2_W},\quad X(x_t)=1.469\pm0.017.
\label{smwc}
\end{align}

New interactions that conserve lepton number and have no new light particles can generate  ${\cal O}_{L,R}^{ij}$.
In contrast, the operators ${\cal O}_{L,R}^{\prime~ij}$ are present when there are light right-handed neutrinos.
Off diagonal (lepton flavour violating) operators occur for example, in models with leptoquarks.
Other possibilities, such as scalar or tensor operators, are not discussed here.

\cref{eq:effHb}  has contributions to $B\to K^{(*)}\nu\bar\nu$   that interfere with the SM, ${\cal O}_{L,R}^{ii}$;
and others that do not, ${\cal O}_{L,R}^{i\neq j}$  and ${\cal O}^{\prime~ij}_{L,R}$.
In $B\to K\nu\bar\nu$ only the vector current enters the hadronic matrix element making the contributions to the rate from
${\cal O}_L^{ij}$ and ${\cal O}_R^{ij}$ (or from ${\cal O}_L^{\prime~ij}$ and ${\cal O}_R^{\prime~ij}$) the same.
In $B\to K^*\nu\bar\nu$ both the vector and axial-vector currents enter the hadronic matrix element resulting in different contributions from
${\cal O}_L^{ij}$ and ${\cal O}_R^{ij}$ as well as from ${\cal O}_L^{\prime~ij}$ and ${\cal O}_R^{\prime~ij}$.
The different neutrino chirality eliminates interference between the contributions from primed and un-primed operators for massless neutrinos. 
Expressions for $R_{K^{(*)}}^{\nu\nu}$ can be determined analytically \cite{Browder:2021hbl} and we find
\begin{subequations}
\begin{align}
R_K^{\nu\nu}  = & 1 
+ {2 C_{L\,\tt SM} \over 3 |C_{L\,\tt SM}|^2}
\sum_i {\rm Re} ( C^{ii}_{L}+C^{ii}_R )
+{1\over 3 |C_{L\,\tt SM}|^2 } \sum_{ij}\left( \left|C^{ij}_L+C^{ij}_R\right|^2 +\left|C_{L}^{\prime~ij}+C_{R}^{\prime~ij}\right|^2 \right),
\\
R_{K^*}^{\nu\nu}  = & 1 
+ {2 C_{L\,\tt SM} \over 3 |C_{L\,\tt SM}|^2}
\sum_i {\rm Re} \left( C^{ii}_{L} \right)
+{1\over 3 |C_{L\,\tt SM}|^2 } 
\sum_{ij}\left( \left|C^{ij}_L\right|^2 +\left|C^{ij}_R\right|^2 +\left|C_{L}^{\prime~ij}\right|^2 +\left|C_{R}^{\prime~ij}\right|^2 \right)
\nonumber
\\
& - 2 \eta \left[
{C_{L\,\tt SM} \over 3 |C_{L\,\tt SM}|^2}
\sum_i {\rm Re} \left( C^{ii}_{R} \right)
+{1\over 3 |C_{L\,\tt SM}|^2 } 
\sum_{ij} {\rm Re}\left(C^{ij}_LC^{*ij}_R+C_{L}^{\prime~ij}C_{R}^{\prime *~ij}\right)
\right], 
\label{eq:scavecbr}
\end{align}
\end{subequations}
where 
\begin{align}
  \eta \equiv {F_- \over F_+}, \,
  F_\pm \equiv \int_0^{(1 -\sqrt{x_K})^2} dx 
  \lambda^{1\over 2}(1, x_K, x) \left[ x A_1^2 + 
  {32\, x_K \over (1 + \sqrt{x_K})^2} A_{12}^2
  \pm  { x \lambda(1, x_K, x) \over (1 + \sqrt{x_K})^4 } V_0^2  \right], 
\end{align}
with $x \equiv q^2/m_B^2$ and $x_K \equiv m_{K^*}^2/m_B^2$.
Numerically, $\eta = 0.63\pm 0.09$ when using the recent light-cone sum rule (LCSR) result for the form factors $A_{1}, A_{12}, V_0$ \cite{Gubernari:2018wyi},
${2 C_{L\,\tt SM} \over 3 |C_{L\,\tt SM}|^2} = -(0.105\pm 0.001)$ and ${1 \over 3 |C_{L\,\tt SM}|^2} = 0.0083\pm 0.0002$.
With these numbers, this expression  agrees with the result presented in \cite{He:2021yoz} that was obtained by using $\tt flavio$ \cite{Straub:2018kue}.
It is interesting to note that the difference $R_K^{\nu\nu} - R_{K^*}^{\nu\nu}$ involves  only the interference terms between the left and right handed currents, 
\begin{align}
R_K^{\nu\nu} - R_{K^*}^{\nu\nu}  =
2(1+\eta)\left[
{C_{L\,\tt SM} \over 3 |C_{L\,\tt SM}|^2}
\sum_i {\rm Re} \left( C^{ii}_{R} \right)
+{1\over 3 |C_{L\,\tt SM}|^2 } 
\sum_{ij} {\rm Re}\left(C^{ij}_LC^{*ij}_R+C_{L}^{\prime~ij}C_{R}^{\prime *~ij}\right) 
\right].
\end{align}

\begin{figure}
\begin{center}
\includegraphics[width=6cm]{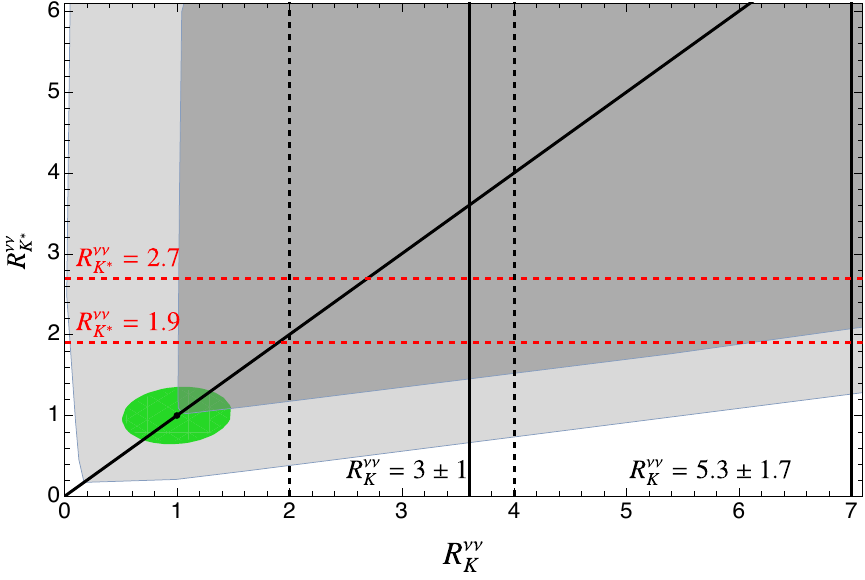}\quad\quad
\includegraphics[width=6cm]{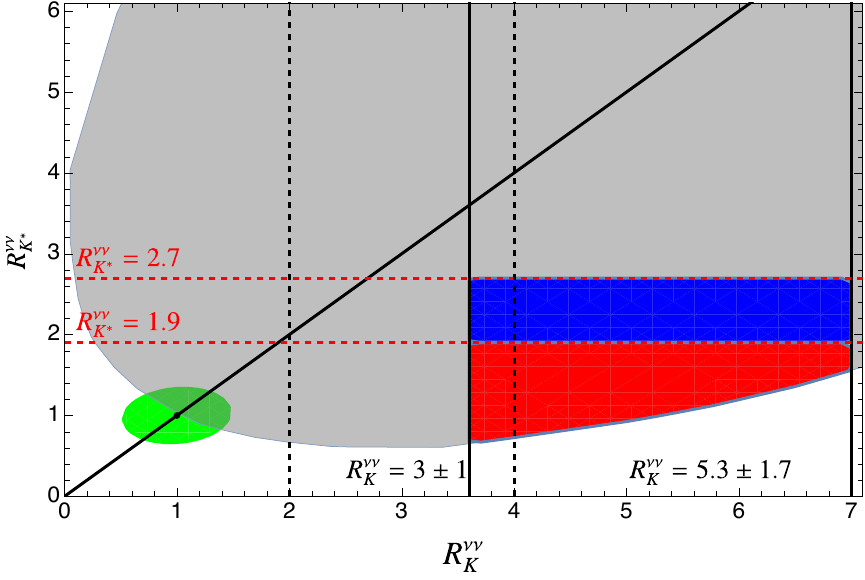}\quad\quad
\vspace{-0.7cm}
\end{center}
\caption{The correlation between $R^{\nu\nu}_{K}$  and $R^{\nu\nu}_{K^*}$  scanning the 12 parameters  $C_L^{ij},C_R^{ij}$ is shown in the left panel as the shaded  light gray region. The corresponding 12 parameter scan in $C_L^{\prime~ ij},C_R^{\prime~ij}$ is shown as the shaded dark gray region. 
The figure highlights in green the $3\,\sigma$  SM prediction. The right panel shows the corresponding scan including only the 6 parameters $C_R^{ij}$. 
The points highlighted in blue or red fall within the new $1\sigma$ range of  $R^{\nu\nu}_K$  (also marked by the vertical solid black lines).
Additionally, the blue (red) points satisfy  $R^{\nu\nu}_{K^*}\leq 2.7 ~(1.9)$ respectively (values marked by the horizontal dashed red lines).
The diagonal black line marks $R_K=R_{K^*}$.}
\label{f:scan}
\end{figure}

We begin by examining the correlations between  $R^{\nu\nu}_{K}$ and $R^{\nu\nu}_{K^*}$ implied by \cref{eq:scavecbr} in light of the new $1\,\sigma$ range.
These are illustrated in \cref{f:scan}  for  a scan of the 12 parameter space  $C_L^{ij},C_R^{ij}$ in light gray in the left panel,
and for the 12 parameter space $C_L^{\prime~ij},C_R^{\prime~ij}$ in darker gray in the left panel\footnote{There are only six  parameters for each of  $C^{(\prime)}_{L,R}$ because the expressions in \cref{eq:scavecbr} are symmetric in $i\leftrightarrow j$. In addition, we assume CP is conserved and use real Wilson coefficients for our scans.}. The primed operators do not interfere with the SM so the points in the darker shaded region satisfy $R^{\nu\nu}_{K^{(*)}}\geq 1$.
In addition, since there is no interference between the primed and unprimed coefficients (we ignore neutrino masses),
it will suffice to look at the two cases separately.
We will focus on  groups that reproduce the new $1\,\sigma$ range of $R^{\nu\nu}_K$,
$3.6\leq R^{\nu\nu}_K\leq 7$, or the new average, $2\leq R^{\nu\nu}_K\leq 4$ and these regions are marked by the vertical solid (dashed) black lines.
We are also interested in groups that satisfy $R^{\nu\nu}_{K^*}\leq 2.7$ or $R^{\nu\nu}_{K^*}\leq 1.9$ and these two regions are bounded  by the horizontal dashed red lines. For comparison we show the three-sigma SM region in green (only theoretical parametric errors as estimated by {\tt flavio}).
The figure shows that it is possible to reproduce values of  $R^{\nu\nu}_{K}$ and $R^{\nu\nu}_{K^*}$ in any of the desired regions with either the 12 unprimed or primed coefficients, albeit with increasing difficulty for the region where  $R^{\nu\nu}_{K}$ matches the new measurement and $R^{\nu\nu}_{K^*}\leq 1.9$. The solid diagonal line, where $R^{\nu\nu}_{K}=R^{\nu\nu}_{K^*}$, will be of interest for certain cases that we discuss below   \cite{He:2021yoz}. 

On the right panel of \cref{f:scan}  we consider only the six $C_R^{ij}$ coefficients,
as these are the ones induced by certain leptoquarks that can satisfy both the
$R^{\nu\nu}_{K}$ and $R^{\nu\nu}_{K^*}$ constraints as  discussed below.
The red and blue regions of interest are the targets for our study and the color code will serve to map predictions for other modes with parameters producing values of  $R^{\nu\nu}_{K}$ and $R^{\nu\nu}_{K^*}$ in the respective region.

To generate the operators with only left-handed neutrinos, ${\cal O}_{L,R}^{ij}$, 
we consider scalar $S$ and vector $V$ leptoquarks coupling to SM fermions as \cite{Davies:1990sc,Davidson:1993qk,Dorsner:2016wpm,He:2021yoz,Lizana:2023kei},
\begin{subequations}
\label{eq:lepcou}
\begin{align}
{\cal L}_S =& \lambda_{LS_0} \bar q^c_L i \tau_2 \ell_L S_0^\dagger
+ \lambda_{L\tilde S_{1/2}} \bar d_R \ell_L \tilde S^\dagger_{1/2}
+ \lambda_{LS_1}\bar q^c_L i\tau_2 \vec \tau \cdot \vec S^\dagger_1 \ell_L + {\rm ~h.c}.,
\\
{\cal L}_V =&  \lambda_{L V_{1/2}} \bar d_R^c \gamma_\mu \ell_L  V^{\dagger \mu}_{1/2}
+ \lambda_{LV_1}\bar q_L \gamma_\mu \vec \tau\cdot  \vec V^{\dagger \mu}_1 \ell_L + {\rm ~h.c.}.
\end{align}
\end{subequations}
The leptoquark fields appearing in \cref{eq:lepcou}, and their transformation properties under the SM group, are
\begin{subequations}
\begin{eqnarray}
S^\dagger_0 &=& S_0^{1/3} : (\bar 3, 1, 1/3), \quad
\tilde S_{1/2}^\dagger = 
\left ( \tilde S_{1/2}^{-1/3}, \tilde S_{1/2}^{2/3} \right ): ( 3, 2,1/6),
\\
\vec \tau\cdot \vec S_1^{\dagger} &=&
\left (
\begin{array}{cc} 
S^{1/3}_1&\sqrt{2} S^{4/3}_1
\\ 
\sqrt{2} S^{-2/3}_1&-S^{1/3}_1
\end{array}
\right ): (\bar 3, 3, 1/3),
\\
V_{1/2}^\dagger &=& 
\left ( V_{1/2}^{1/3}, V_{1/2}^{4/3} \right ): ( \bar 3, 2, 5/6),\quad
\vec \tau\cdot \vec V_1^{\dagger}  =
\left (
\begin{array}{cc} 
V^{2/3}_1&\sqrt{2} V^{5/3}_1
\\ 
\sqrt{2} V^{-1/3}_1&-V^{2/3}_1
\end{array}
\right ): ( 3, 3, 2/3).
\end{eqnarray}
\end{subequations}

Exchange of $S_0,S_1$ or $V_1$, generates the $C_L$ coefficients only,
whereas exchange of $\tilde{S}_{1/2}$ or $V_{1/2}$ generates only the $C_R$ coefficients.
We find that  it is not possible to satisfy the bound on $R^{\nu\nu}_{K^*}$, \cref{eq:rkstar},
with only $C_L^{ij}$ coefficients because they predict $R_{K}^{\nu\nu}=R_{K^*}^{\nu\nu}$
and thus fall along the diagonal line shown in the left panel of \cref{f:scan}.
Models with only  $S_0,S_1$ or $V_1$ are thus incompatible with the new Belle II measurement.
The same figure shows that they could satisfy the new average value of $R^{\nu\nu}_{K}$ but would only satisfy the weaker bound on $R^{\nu\nu}_{K^*}$. 

These three leptoquarks are also constrained by modes with charged leptons. 
$S_0$ induces charged current operators and contributes to the ``charged $B$ anomalies'' in $R_{D^{(*)}}$. 
If we choose its parameters to reach the largest point consistent with the new average value of
$R^{\nu\nu}_{K}$ while satisfying the looser constraint on $R^{\nu\nu}_{K^*}$, $R_{K}^{\nu\nu}=R_{K*}^{\nu\nu}\sim 2.7$,
then we find that it can only enhance $R_{D^{(*)}}$ by about 6\% over their SM value.
This is insufficient to explain the current discrepancy as reported by HFLAV,
$r_{D}=1.19\pm0.10,\, r_{D^{*}}=1.12\pm 0.06 $ \cite{PhysRevD.107.052008}, $r_{D^{(*)}}=R_{D^{(*)}}/(R_{D^{(*)}})_{\rm SM}$.
Conversely, tuning the parameters of $S_0$ to $r_{D^{(*)}}=1.12$ results in $R_{K}^{\nu\nu}=R_{K*}^{\nu\nu}\gtrsim 14$.

The cases of $S_1$ and $V_1$ are also constrained by the ``neutral $B$ anomalies'' as they generate the correlated coefficients \cite{He:2021yoz}
\begin{subequations}
\begin{align}
S_1 & :~~ C_9^{ij}=-C_{10}^{ij}= 2 C_L^{ij},
\\
V_1 & :~~ C_9^{ij}=-C_{10}^{ij}= \frac{1}{2} C_L^{ij}.
\end{align}
\end{subequations}
Recent global fits suggest that the observables in $b\to s\ell^+\ell^-$ are best described with
$C_9^{\mu\mu}\sim C_9^{ee}\sim -1$ implying  $R^{\nu\nu}_{K^{}}=R^{\nu\nu}_{K^{*}} \lesssim 1.1$ for
$S_1$ and $R^\nu_{K^{(*)}}\lesssim 1.5$ for $V_1$.

For the case where new physics appears only with  $C_R$ coefficients, on the other hand, 
there is a large region of parameter space that simultaneously satisfies the new $1\,\sigma$ range of $R^{\nu\nu}_K$
and the 90\,\% C.L. upper bound on $R^{\nu\nu}_{K^*}$ as seen on the right panel of \cref{f:scan}.
However, if we restrict ourselves to models where only the off-diagonal (in lepton flavour) coefficients,
$C_R^{i\neq j}$, are not zero, it is impossible to satisfy both constraints simultaneously because this scenario also results in $R^{\nu\nu}_K=R^{\nu\nu}_{K^*}$.  

We are left with two leptoquarks that can satisfy both  the new $1\,\sigma$ range of $R^{\nu\nu}_K$ and the 90\,\% C.L. upper bounds on $R^{\nu\nu}_{K^*}$.
They are $\tilde{S}_{1/2}$ and $V_{1/2}$ with both diagonal and off diagonal flavour couplings. They generate the coefficients \cite{He:2021yoz}
\begin{align}
C_R^{ ij}& =C_{9^\prime}^{ij} = - C_{10^\prime }^{ij} 
= \frac{\pi}{ \sqrt{2} \alpha G_F V_{tb}V_{ts}^*}\left (- \frac{\lambda^{2j}_{L\tilde S_{1/2}} \lambda^{* 3i}_{L\tilde S_{1/2}}}{ 2 m_{S_{1/2}}^2}
+ \frac{\lambda^{3j}_{L V_{1/2}} \lambda^{* 2i}_{L V_{1/2}}}{ m_{V_{1/2}}^2}
\right),
\label{eq:wcflq}
\end{align}
and are thus correlated with other modes with charged leptons \cite{Bause:2021cna,Rajeev:2021ntt}. 
The flavour diagonal entries for the first two generations affect the $b\to s \ell^+\ell^-$ processes.
Recent global fits\footnote{After the corrected LHCb value of $R_{K(K^*)}$\cite{LHCb:2022qnv}, 
for a review with references see \cite{Capdevila:2023yhq}.} to  $b\to s \ell^+\ell^-$ observables, 
in addition to preferring  $C_9^{\mu\mu}\sim C_9^{ee}\sim -1$, also allow $C_{9^\prime}^{ii}$ and $C_{10^\prime}^{ii}$ to be non-zero,
but much smaller than one. This makes it obvious that  $\tilde{S}_{1/2}$ and $V_{1/2}$ are not  
a preferred solution for the ``neutral $B$ anomalies''.  
$\tilde{S}_{1/2}$ and $V_{1/2}$ can still modify $R^{\nu\nu}_K$ and $R^{\nu\nu}_{K^*}$ 
but they are constrained to produce small values of $C_{9^\prime}^{\mu\mu,ee}$ and 
$C_{10^\prime}^{\mu\mu,ee}$ and additional NP would be needed to provide the preferred values of $C_9$.
 
\begin{figure}
\begin{center}
\includegraphics[width=5cm]{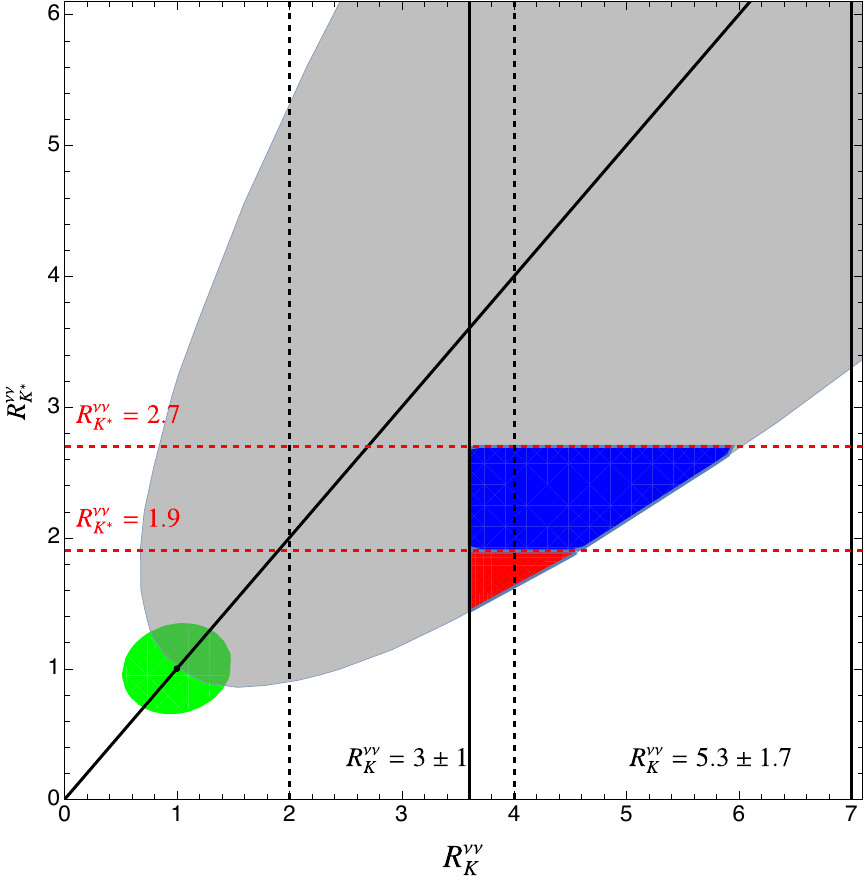}\quad
\includegraphics[width=5cm]{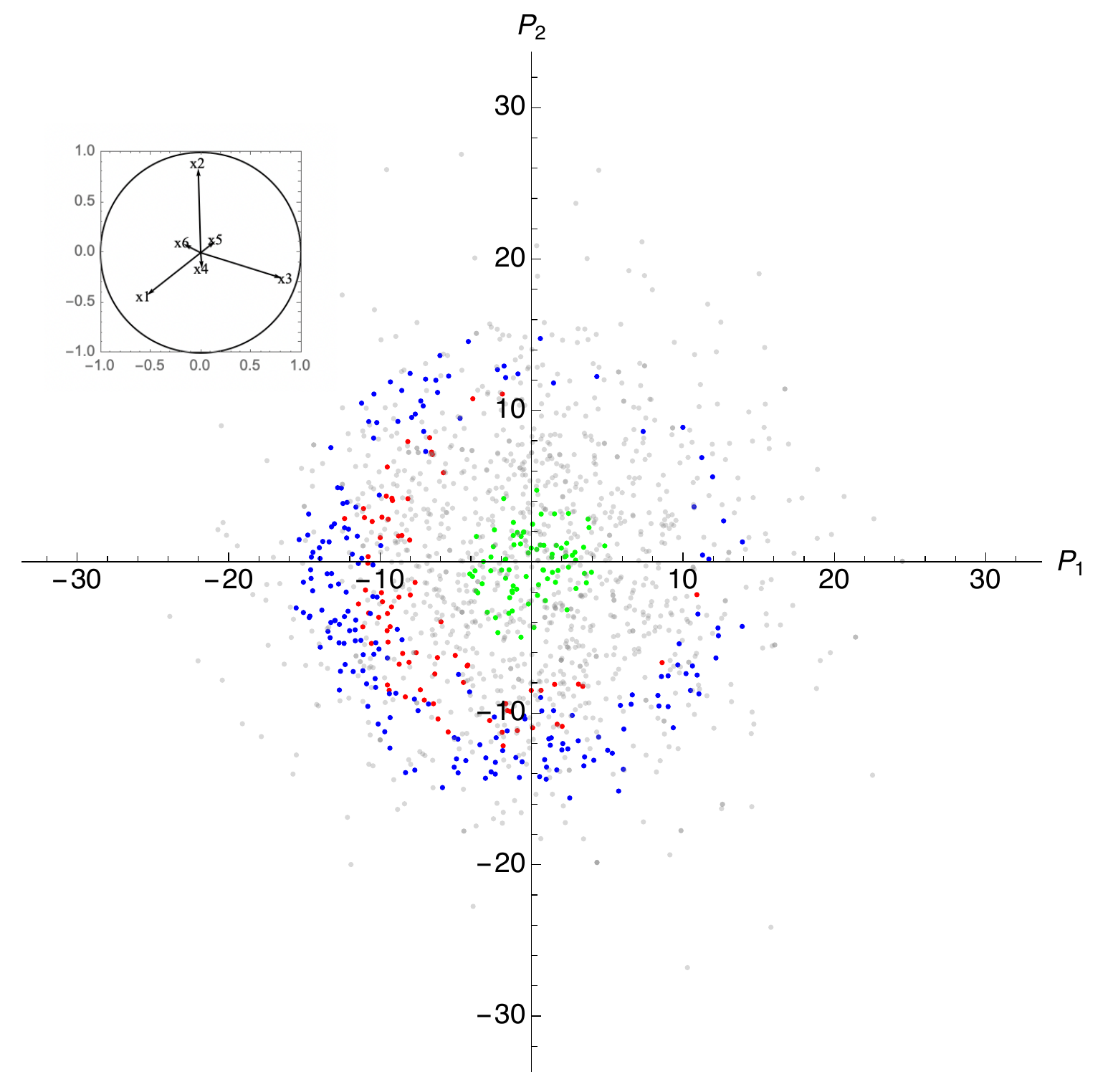}\quad
\includegraphics[width=5cm]{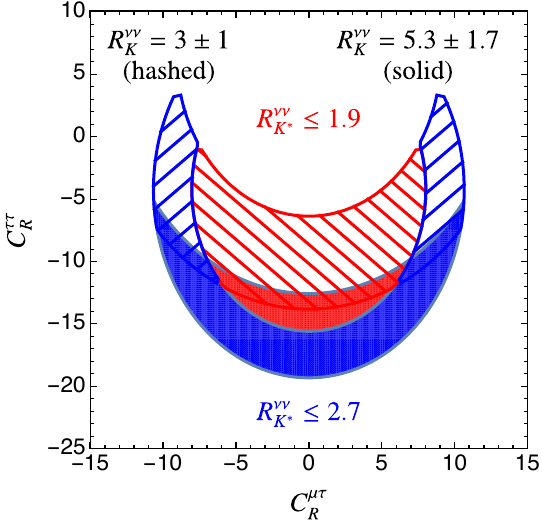}
\vspace{-0.7cm}
\end{center}
\caption{Left panel: scan of parameters $C_R^{ij}$ where the only non-zero diagonal term is $C_R^{\tau\tau}$ with the same color code as in \cref{f:scan}.
Centre  panel: a selected slice of a projection in parameter space illustrating the location of the red and blue clusters in the six-parameter space of $C_R^{ij}$. 
As this figure suggests, these clusters concentrate in regions where at least one of the diagonal entries $|C_R^{ii}|$ is near 10.
Right panel: parameter region allowed when only $C_R^{\tau\tau}$ and $C_R^{\mu\tau}$ are non-zero,
we have added in this panel the hashed regions showing the corresponding solutions when the average value of $R^{\nu\nu}_K$ is used.}
\label{f:lqraa}
\end{figure}

A scan of the six-parameter space for a symmetric $C_R^{ij}$ is shown on the left panel of \cref{f:lqraa},
marking the regions that satisfy \cref{eq:newrn} and \cref{eq:rkstar} as before.
With the aid of tools described in \cite{Laa:2019bap,Laa:2021dlg,laa2022new} we can examine the parameter region responsible for the blue and red regions.
We show in the centre panel a slice of a projection mostly onto the diagonal elements $C_R^{ii}$ (axes labeled $x_{1,2,3}$).
The slice is thin in the orthogonal space, mostly the off-diagonal elements  $C_R^{i\neq j}$.
The inset represents the projection matrix corresponding to this view. 
The visualisation indicates that the red and blue points concentrate in regions where at least one diagonal $C_R^{ii}$ element is large,
${\cal O}(10)$. Given that the global fits $b\to s \ell^+\ell^-$ do not admit solutions with large $C_{9^\prime,10^\prime}^{ee,\mu\mu}$
we conclude that the case with large $C_R^{\tau\tau}$ is the only viable solution.\footnote{A short video illustrating the tools arriving at the centre panel of \cref{f:lqraa} can be found in the supplemental material on ArXiv.} 
For further illustration, we show the viable two-parameter space $C_R^{\mu\tau}-C_R^{\tau\tau}$ region on the right panel of the same figure
(the colour code is the same throughout).

Quantitatively, in this case, we find a lower bound of $R^{\nu\nu}_{K^*}\geq 1.5$ for the red region.
It corresponds to  $C_R^{\mu\tau}\sim C_R^{\tau\mu}\sim 0$ and $C_R^{\tau\tau}\sim -13.2$.
Similar solutions exist for small but non-zero $C_R^{ee},C_R^{\mu\mu}\sim{\cal O}(0.1)$ in combination with a large $C_R^{\tau\tau}\sim {\cal O}(10)$. 

\begin{figure}
\begin{center}
\includegraphics[width=5cm]{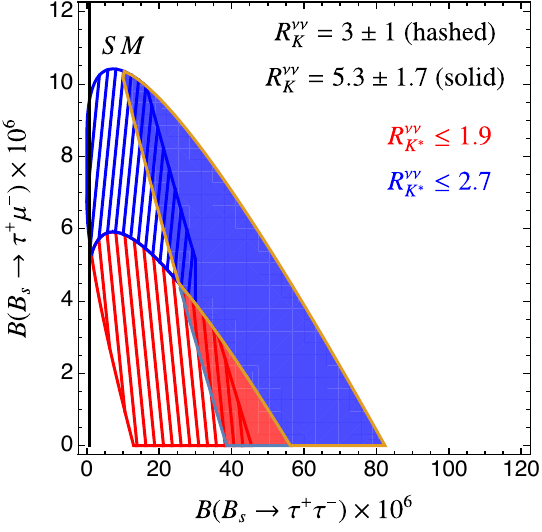}\quad\quad
\includegraphics[width=5cm]{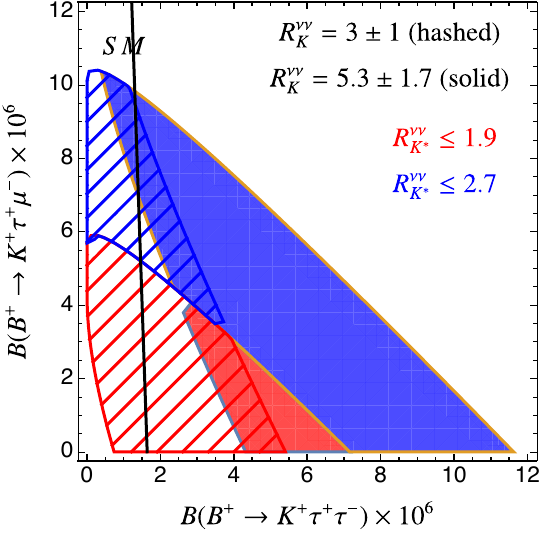}
\vspace{-0.7cm}
\end{center}
\caption{Mapping of the parameter space selected in \cref{f:lqraa} into predictions for other $B$ meson decay modes.
The solid black line for the lepton flavour conserving processes marks the SM prediction.} 
\label{f:lq33impl}
\end{figure}

The Wilson coefficients generated by either one $\tilde{S}_{1/2}$ or $V_{1/2}$, \cref{eq:wcflq},
imply large rates for other $B$ decay modes when chosen to produce values of  $R^{\nu\nu}_{K^{(*)}}$ in the red/blue regions as we illustrate in \cref{f:lq33impl}.
We see that the CLFV branching fractions ${\cal B}(B^+\to K^+\mu^-\tau^+)$ and ${\cal B}(B_s\to \mu^-\tau^+)$ can reach values of $10^{-5}$,
which are within factors of two below their current experimental upper limits:
$2.8 \times 10^{-5} (90\,\%)$ and $4.2 \times 10^{-5} (95\,\%)$ \cite{ParticleDataGroup:2022pth} respectively.
Similarly, the branching ratios for the lepton flavour conserving modes ${\cal B}(B^+\to K^+\tau^+\tau^-)$
and ${\cal B}(B_s\to \tau^+\tau^-)$ can reach values of a few times $10^{-5}$,
whereas their current experimental upper limits are  ${\cal B}(B^+\to K^+\tau^+\tau^-)\leq 2.25 \times 10^{-3} (90\,\%)$ and ${\cal B}(B_s\to \tau^+\tau^-)\leq 6.8 \times 10^{-3} (95\,\%)$ \cite{ParticleDataGroup:2022pth}. 
These represent large enhancements over the SM (shown as a vertical black line in \cref{f:lq33impl}) and could rule out this possibility in the future. 

The case of leptoquarks was also discussed by \cite{Bause:2023mfe}.\footnote{Which appeared as we were finalizing our manuscript.}
They reach similar but not identical conclusions to us. 
For $S_0,S_1$ or $V_1$ (called $S_1,S_3$ and $V_3$ in that reference) 
we agree that only $C_L$ coefficients are generated and this results in $R^{\nu\nu}_{K}=R^{\nu\nu}_{K^{*}}$. 
Our \cref{f:scan} quantifies just how much this deviates from the current measurements. 
We also agree that global fits to $b\to s\ell^+\ell^-$ processes imply that significant deviations from the SM would correlate with $\tau$-lepton flavour.
We disagree with the conclusions reached for the case of $\tilde{S}_{1/2}$ or $V_{1/2}$ (called $\tilde{S}_2,V_2$ in \cite{Bause:2023mfe}),
which induce only $C_R$. 
We find that these two can explain the measurements of $R^{\nu\nu}_{K}$ and $R^{\nu\nu}_{K^{*}}$
as shown in \cref{f:lqraa} for a certain region of parameter space, contrary to what is claimed in \cite{Bause:2023mfe}.

We briefly turn our attention to models with a light sterile neutrino (that couples to SM fields via a $Z^\prime$) \cite{He:2002ha,He:2017bft}.
This model produces non-zero  $C_L^{\prime~ij},C_R^{\prime~ij}$ as those in the left panel of \cref{f:scan}.
In this case, \cref{eq:scavecbr} implies that in the region where  $3.6\leq R^{\nu\nu}_K\leq 7$ there is a lower bound $R^{\nu\nu}_{K^*}\geq 1.47$.
Both  $C_L^{\prime~ij}\neq 0$ and $C_R^{\prime~ij}\neq 0$ are needed to deviate from  $R^{\nu\nu}_{K}=R^{\nu\nu}_{K^*}$
and the parameter scan shows that there could be solutions of this type, 
the region satisfying both $R^{\nu\nu}_{K},~R^{\nu\nu}_{K^*}$ constraints is shown in dark gray in \cref{f:scan}.
However, the specific model described in \cite{He:2021yoz} is severely constrained by both $Z-Z^\prime$ mixing and by $B_s$ mixing \cite{He:2004it,He:2006bk}.
When we  consider these two constraints, the model predicts  $R^{\nu\nu}_{K}\approx R^{\nu\nu}_{K^*}\lesssim 2$,
barely reaching the lower end of the new $R^{\nu\nu}_{K}$ average.

\section{Light dark matter scenario for $B\to K + invisible$}

A large  ${\cal B}(B^+ \to K^+ \nu \bar \nu)$ as the one reported by Belle II can also be caused by invisible particles beyond the SM.
In this case  light dark matter (DM) pairs take the place of the neutrinos,
and we now examine this possibility.\footnote{Similar considerations for charm decay can be found in \cite{Fajfer:2021woc, Li:2023sjf}.} 
Since the spin of the possible DM particles is not known, we consider three cases: a spin-0 scalar $\phi$,
a spin-1/2 fermion $\chi$, or a spin-1 vector $X$, respectively. 
We work within the framework of LEFT \cite{Jenkins:2017jig,Liao:2020zyx},
so we only impose the SM broken phase symmetry $SU(3)_{\rm c}\times U(1)_{\rm em}$ on the effective operators.
In this way, the LEFT framework  covers  scenarios that contain both light DM as well as new weak scale mediators that have been integrated out.
The complete  LEFT operator basis with a pair of light DM fields at leading order was recently provided by us in Ref.\,\cite{He:2022ljo}.
In the following, we will adopt the notation in that paper and list the relevant FCNC local quark-DM operators
mediating $B^+ \to K^+ +invisible$ decays for all cases. 

For this discussion, it is not convenient to use the ratio $R^K_{\nu\nu}$ of \cref{eq:newm}.
Instead, it is better to look at the difference between the measurement and the SM, and refer to it as the ``NP window''.
Using the new Belle II measurement this is,
{\small
\begin{align}
\label{eq:np_window}
{\cal B}(B^+ \to K^+ +invisible)_{\tt NP}
\equiv   
{\cal B}(B^+ \to K^+\nu\bar\nu)_{\tt exp}
- {\cal B}(B^+ \to K^+\nu\bar\nu)_{\tt SM} 
= (1.9\pm 0.7)\times 10^{-5},
\end{align} }
and it becomes $(1.0\pm 0.4)\times 10^{-5}$ using instead the new average experimental value.

In keeping with our previous discussion, we explore the possibility of attributing the new Belle II result
to light dark particles while satisfying other constraints.
We thus illustrate the implications of requiring the new contribution to the rate $B^+ \to K^+ +invisible$
to fall within the $1\,\sigma$ range of \cref{eq:np_window}.
At the same time, the effective operators studied below will also contribute to the decays
$B^0 \to K^0+invisible$ and $B^{+(0)} \to K^{*+(0)}+invisible$.
There exist experimental upper bounds on these modes (at $90\,\%$ C.L.) that we will use to constrain the parameter space.
A final consideration relevant to this scenario is the observation by Belle II that the excess of events
is predominantly concentrated in the region with $3\leq q^2\leq7$ GeV$^2$ \cite{Belle-II:2023esi}.   

It has been pointed out recently that the excess over the SM from the Belle II measurement
is model dependent \cite{Fridell:2023ssf,Gartner:2024muk}.
A recasting of the Belle II analysis for the different types of particles, different masses,
and different operators that we discuss below is beyond the scope of the present study.
Each choice of particle, mass or operator leads to a different kinematic $(q^2)$ distribution
and this can be exploited by future experimental analysis to differentiate between benchmarks.
In this study we rely on the branching ratio under the assumption that the number reported by Belle II is model independent.
Clearly this is not exactly true, but a reliable model independent branching ratio
can only be extracted by the experimental collaboration (if at all).
Our results should be interpreted as a comparative guide to different scenarios
to select preferred benchmarks that can be studied further if the excess persists.

\subsection{Scalar DM case}

We begin with the spin-0 scalar DM case. The leading order operators mediating $B^+ \to K^+\phi\phi$
consist of quark scalar and vector currents with $s,b$ flavors coupled to two scalar fields,
\begin{align}
\calO_{q\phi}^{S,sb} &=  (\overline{s} b)(\phi^\dagger \phi), 
&
\calO_{q\phi}^{V,sb} &=  (\overline{s}\gamma^\mu b) (\phi^\dagger i \overleftrightarrow{\partial_\mu} \phi). \, (\times) 
\label{eq:Oqphi}
\end{align}
The symbol ``$(\times) $'' indicates an operator that vanishes for real scalar fields,
and $\phi^\dagger\overleftrightarrow{\partial_\mu} \phi \equiv \phi^\dagger (\partial_\mu \phi) - (\partial_\mu \phi^\dagger) \phi$.
These operators occur at dimension six in the SMEFT framework \cite{Kamenik:2011vy}.
In addition to $B^+ \to K^+ \phi\phi$, the scalar current operator also induces the decay $B^{0} \to K^0 \phi\phi$.
The vector operator can also mediate the modes  $B^{0} \to K^{*0}\phi\phi$ and $B^+ \to K^{*+}\phi\phi$. 
The differential decay widths for these four modes have been given in \cite{He:2022ljo}.
For the  hadronic form factors involved, we use the recent lattice calculation in \cite{Parrott:2022rgu} for $B\to K$,
and those from LCSR \cite{Gubernari:2018wyi} for $B\to K^*$. 
To quantify the allowed parameter space, we define an effective heavy scale associated with each operator as
$C_{q\phi}^{S,sb} \equiv \Lambda_{\rm eff}^{-1}$ and $C_{q\phi}^{V,sb} \equiv \Lambda_{\rm eff}^{-2}$. 

\begin{figure}
\begin{center}
\includegraphics[width=5cm]{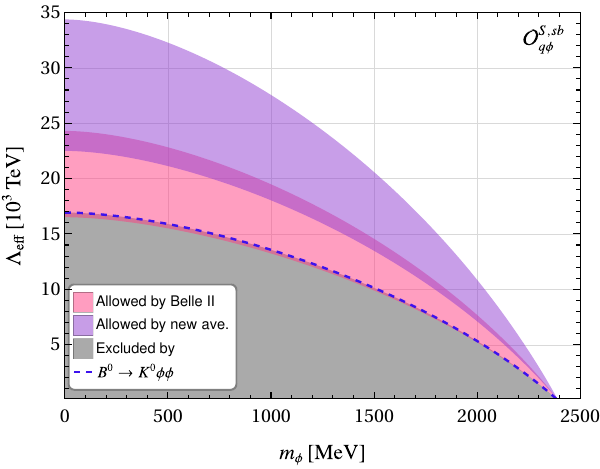}\quad\quad
\includegraphics[width=5cm]{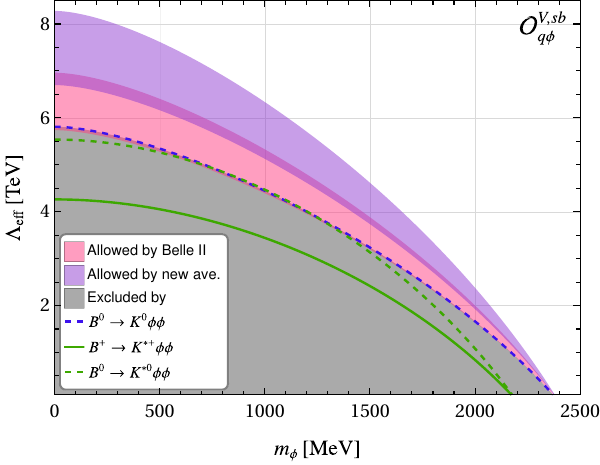}
\vspace{-0.7cm}
\end{center}
\caption{The pink [purple] region is the parameter space that could explain the recent Belle II excess [new average]
with scalar DM for scalar (vector) quark current operators ${\cal O}_{q\phi}^{S,sb} ({\cal O}_{q\phi}^{V,sb})$.
The gray region is excluded by other $B$ meson decay modes that are indicated in the plots by colour lines. }
\label{fig:scalar_limit}
\end{figure}

\cref{fig:scalar_limit} shows our results for the operators in \cref{eq:Oqphi}.
The green (solid and dashed) lines and the blue dashed line mark the lower limit
on the scale $\Lambda_{\rm eff}$ associated with the scalar (vector) operator
in the left (right) panel resulting from the upper bound on $B^{+(0)} \to K^{*+(0)}\phi\phi$ and $B^{0} \to K^{0}\phi\phi$ respectively.
The pink region covers the parameter space in the $m_\phi-\Lambda_{\rm eff}$ plane that  accommodates the Belle II $1\,\sigma$ range.
The purple region indicates the allowed parameter space once the new average number is adopted,
the thin dark purple strip is the overlapping region.
The gray area is excluded by the experimental upper limits and is mostly due to the neutral modes $B^0 \to K^{(*)0}\nu\bar\nu$ for these two operators.
The scalar operator provides a larger parameter region consistent with the new Belle II measurement than the vector operator.
The vector operator, however, results in a $q^2$ distribution more in tune with the reported excess, as we illustrate in \cref{fig:dGds_dis}.

\subsection{Fermion DM case}

We now turn to  fermion DM particles.\footnote{A general study of the three-body decay with sterile neutrinos
in the $\nu$SMEFT framework \cite{Liao:2016qyd} can be found in \cite{Felkl:2021uxi,Felkl:2023ayn}.}
Denoting the DM field by $\chi$, there are six leading, dimension six operators
that are responsible for the $B^+ \to K^+\chi\chi$ transition \cite{He:2022ljo},
\begin{subequations}
\label{eq:fer_DM_ope}
\begin{align}
\calO_{q\chi1}^{S,sb} &= (\overline{s} b)(\overline{\chi}\chi),
&
\calO_{q\chi2}^{S,sb} &= (\overline{s} b)(\overline{\chi}i \gamma_5\chi), 
\\
\calO_{q\chi1}^{V,sb} &=  (\overline{s}\gamma^\mu  b)(\overline{\chi}\gamma_\mu  \chi),
\, (\times)
&
\calO_{q\chi2}^{V,sb} &= (\overline{s}\gamma^\mu b)(\overline{\chi}\gamma_\mu  \gamma_5\chi), 
\\
\calO_{q\chi1}^{T,sb} &=  (\overline{s}\sigma^{\mu\nu}  b)(\overline{\chi}\sigma_{\mu\nu}   \chi),
\, (\times)
&
\calO_{q\chi2}^{T,sb} &= (\overline{s}\sigma^{\mu\nu} b)(\overline{\chi}\sigma_{\mu\nu}  \gamma_5\chi), 
\, (\times)
\end{align}
\end{subequations}
where the ``$(\times) $'' indicates the accompanying operator vanishes for Majorana fermions.
The implications for $B\to K$ transitions from such operators have been partially considered before \cite{Li:2020dpc}. 
In Ref.\,\cite{Li:2020dpc}, only the charged processes $B^+\to K^{(*)+}\chi\chi$ are considered
and this was done before the new Belle II measurement. 
Here we use the new experimental result and include all possible modes including the neutral ones.
We also use the new lattice form factors \cite{Parrott:2022rgu}.

\begin{figure}
\begin{center}
\includegraphics[width=5cm]{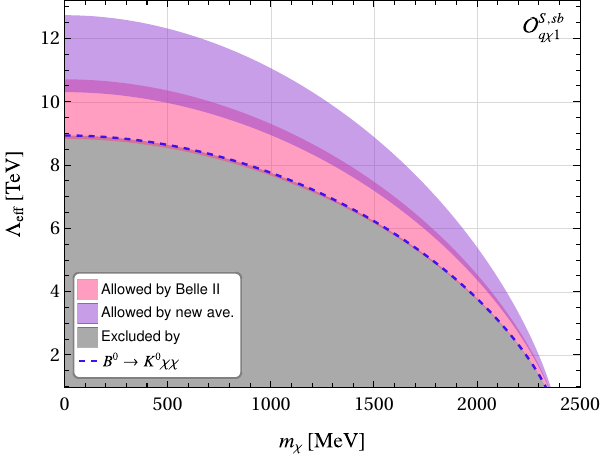} \quad
\includegraphics[width=5cm]{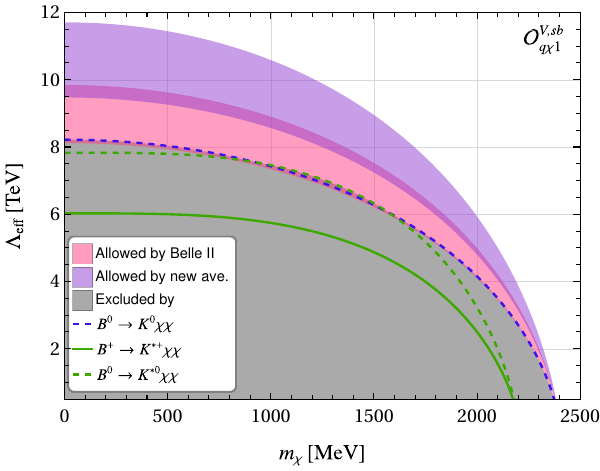} \quad
\includegraphics[width=5cm]{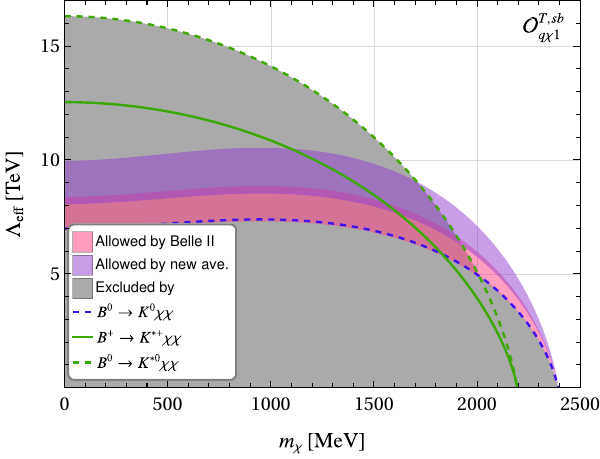}
\vspace{0.2cm}
\\ 
\includegraphics[width=5cm]{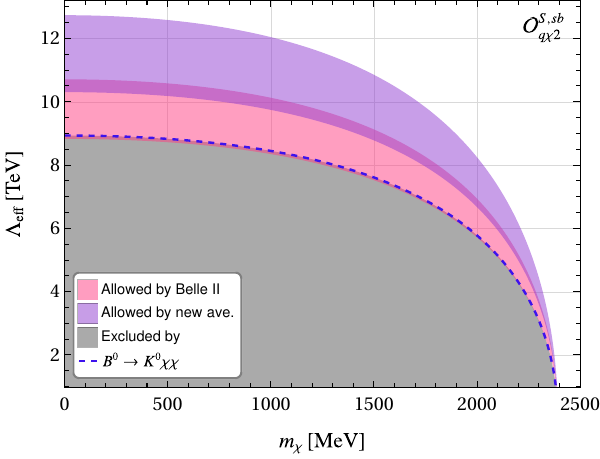} \quad
\includegraphics[width=5cm]{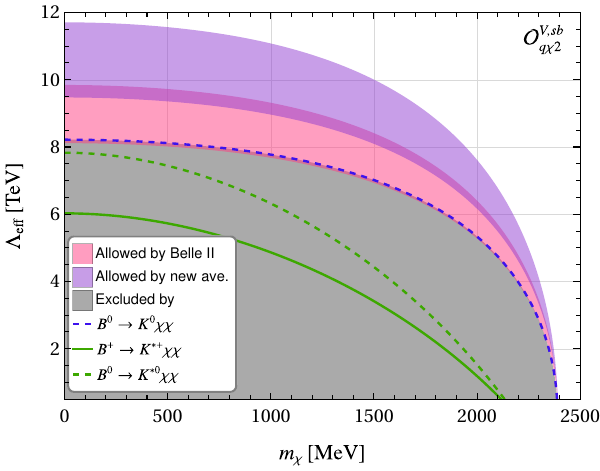} \quad
\includegraphics[width=5cm]{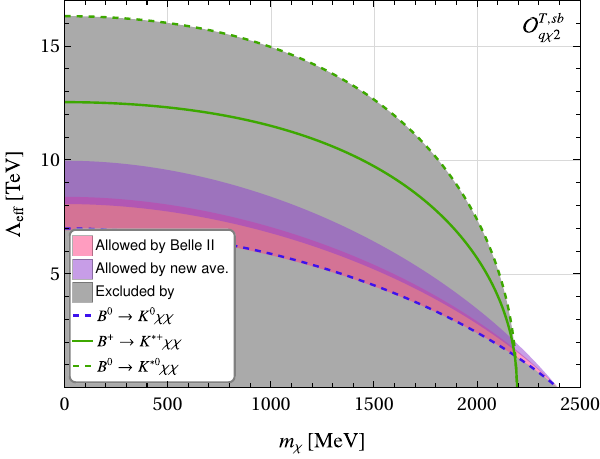}
\vspace{-0.7cm}
\end{center}
\caption{The pink [purple] region shows the parameter space that could explain the recent Belle II excess [new average]
with fermion DM for the operators in \cref{eq:fer_DM_ope}.
The gray region is excluded by other $B$ meson decay modes that are indicated in the plots by coloured lines.}
\label{fig:fermion_limit}
\end{figure}

The differential decay widths for this case have been given previously in \cite{Kamenik:2011vy},
and we reproduce them in terms of our convention in Appendix A for reference.
To study the parameter space, we rewrite the Wilson coefficient $C_{i}^j$ of the operator ${\cal O}_{i}^j$
as an effective scale $C_{i}^j \equiv \Lambda_{\rm eff}^{-2}$. 
For each operator in \cref{eq:fer_DM_ope}, \cref{fig:fermion_limit} shows the parameter space in the $m_\chi-\Lambda_{\rm eff}$ plane
that reproduces the new Belle II $1\,\sigma$  (new average) result as the pink (purple) shaded region.
The gray region is excluded by other $B$ meson decay modes. 
It can be seen that the two operators with tensor quark currents have almost no viable region to explain the excess,
unlike the operators with scalar or vector currents.
Most of the exclusion for these four operators arises from the upper limit on the neutral mode $B^0 \to K^0 \chi\chi$.
If we insist on an excess of events concentrated in the $3\leq q^2\leq 7$~GeV$^2$ bins, only the 
vector current operators ${\cal O}_{q\chi1,2}^{V,sb}$ with certain DM masses remain viable as we illustrate in \cref{fig:dGds_dis}.

\subsection{Vector DM case} 

Finally, we consider the vector DM. There are two parametrizations that can be used in this case as discussed in \cite{He:2022ljo}.
Here we adopt the one with a four-vector field $X_\mu$ for simplicity. The operators have been classified by us in \cite{He:2022ljo},
and the  ones relevant for $B^+ \to K^+XX$ transitions are, 
\begin{subequations}
\label{eq:OqX}
\begin{align}
\calO_{q X}^{S,sb} &=  (\overline{s} b)(X_\mu^\dagger X^\mu), 
\\
\calO_{q X1}^{T,sb} &=   \frac{i}{2} (\overline{s}  \sigma^{\mu\nu} b) (X_\mu^\dagger X_\nu - X_\nu^\dagger X_\mu),  \, (\times) 
\\
\calO_{q X2}^{T,sb} &=  \frac{1}{2} (\overline{s}\sigma^{\mu\nu}\gamma_5 b) (X_\mu^\dagger X_\nu - X_\nu^\dagger X_\mu),  \, (\times) 
\\
\calO_{q X2}^{V,sb} &=  (\overline{s}\gamma_\mu b)\partial_\nu (X^{\mu \dagger} X^\nu + X^{\nu \dagger} X^\mu  ), 
\\
\calO_{q X3}^{V,sb} &=  (\overline{s}\gamma_\mu b)( X_\rho^\dagger \overleftrightarrow{\partial_\nu} X_\sigma )\epsilon^{\mu\nu\rho\sigma}, 
\\
\calO_{q X4}^{V,sb} &= (\overline{s}\gamma^\mu b)(X_\nu^\dagger  i \overleftrightarrow{\partial_\mu} X^\nu), 
 \, (\times)
 \\
\calO_{q X5}^{V,sb} &= (\overline{s}\gamma_\mu b)i\partial_\nu (X^{\mu \dagger} X^\nu - X^{\nu \dagger} X^\mu  ),  \, (\times)
 \\
\calO_{q X6}^{V,sb} &= (\overline{s}\gamma_\mu b) i \partial_\nu ( X^\dagger_\rho X_\sigma )\epsilon^{\mu\nu\rho\sigma}.
\, (\times)  
\end{align}
\end{subequations}
The symbol ``$(\times)$'' indicates that the corresponding operator vanishes for real vector fields.
To address the well-known singularity problem that affects vector fields in the limit of vanishing mass,
for our numerical analysis, we scale the Wilson coefficients of these operators in the following manner, 
\begin{eqnarray}
C_{qX}^{S} \equiv {m^2 \over \Lambda_{\rm eff}^3}, 
\quad
C_{qX1,2}^{T} \equiv {m^2 \over \Lambda_{\rm eff}^3}, 
\quad
C_{qX2,4,5}^{V} \equiv {m^2 \over \Lambda_{\rm eff}^4},
\quad
C_{qX3,6}^{V} \equiv  {m \over \Lambda_{\rm eff}^3}.
\label{eq:CqX2mLam}
\end{eqnarray}

\begin{figure}
\begin{center}
\includegraphics[width=4cm]{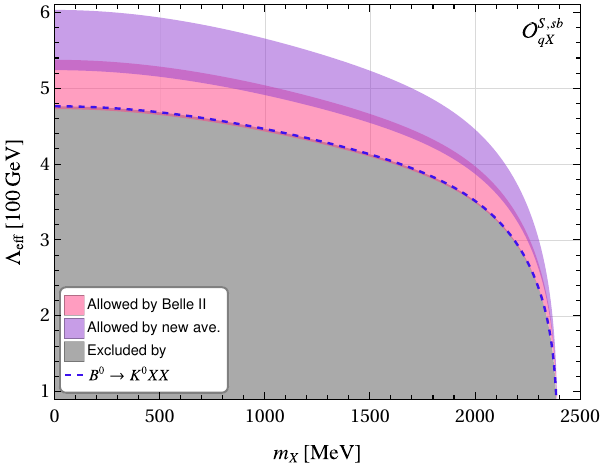}
 \includegraphics[width=4.1cm]{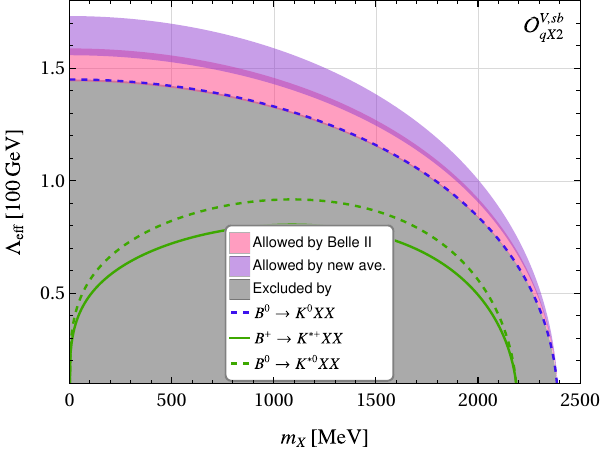}
\includegraphics[width=4.1cm]{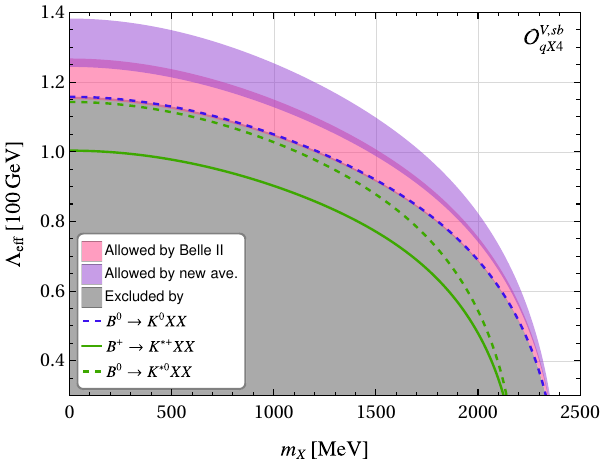} 
\includegraphics[width=4.1cm]{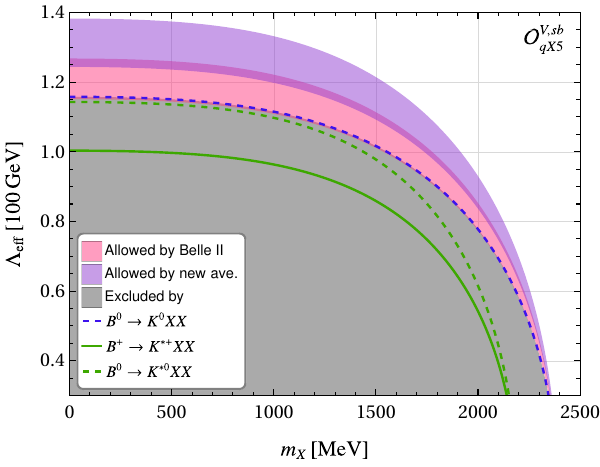}
\vspace{0.2cm}
\\
\includegraphics[width=4cm]{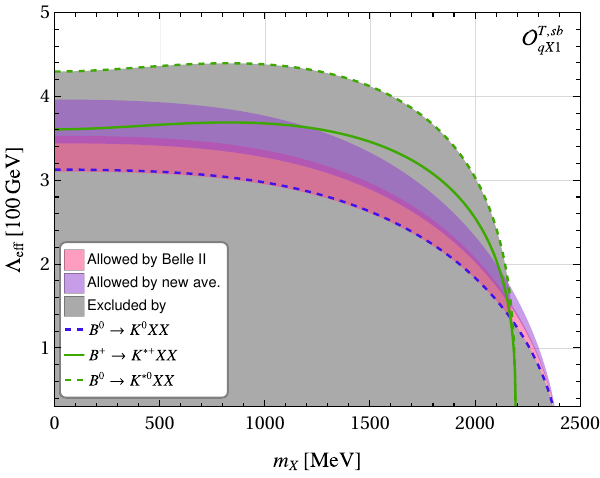}\,
\includegraphics[width=4cm]{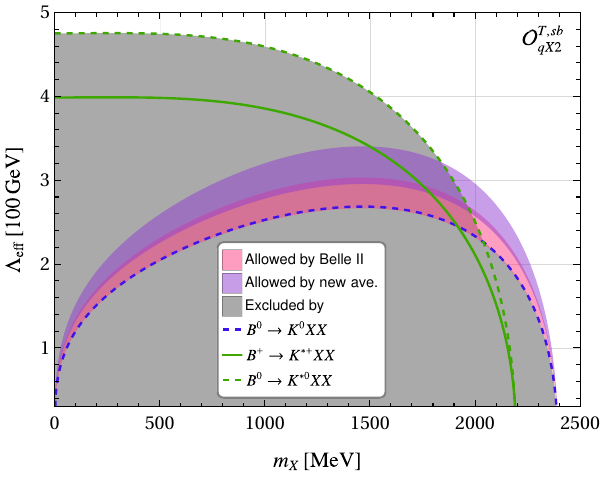}\,
\includegraphics[width=4cm]{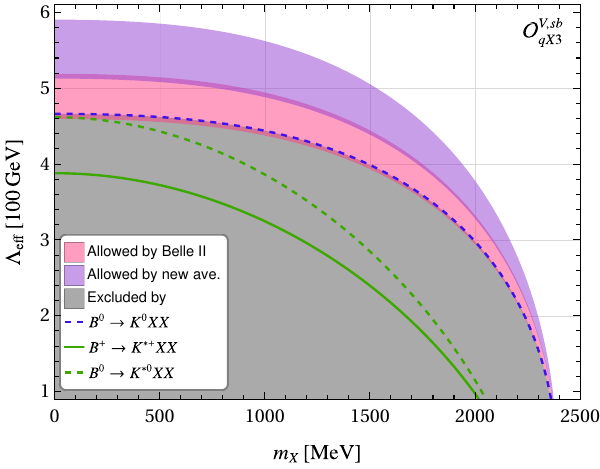} \,
\includegraphics[width=4cm]{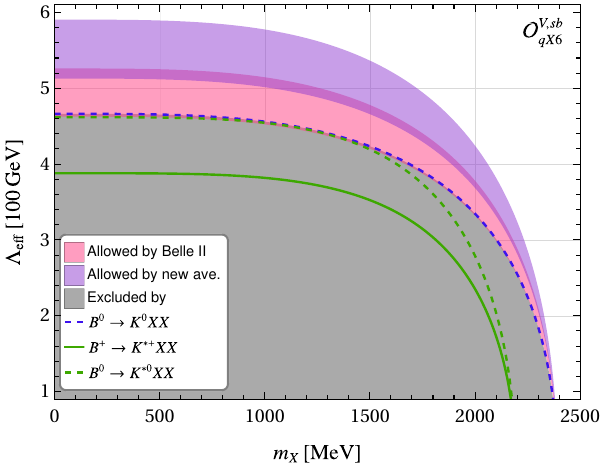}
\vspace{-0.7cm}
\end{center}
\caption{The pink [purple] region shows the parameter space that could explain the recent Belle II excess [new average]
with vector DM for the operators in \cref{eq:OqX}.
The gray region is excluded by other $B$ meson decay modes that are indicated in the plots by coloured lines. }
\label{fig:vector_limit}
\end{figure}

In \cref{fig:vector_limit}, we show the parameter space resulting in a branching ratio in agreement with \cref{eq:np_window}.
It can be seen, that except for the two operators with tensor quark currents ${\cal O}_{qX1,2}^{T,sb}$,
the remaining operators contain a large acceptable parameter region.  

\subsection{The $q^2$ distribution}

The excess of events observed by Belle II appears to occur mainly for $q^2$ values
between $3-7\,\rm GeV^2$ \cite{Belle-II:2023esi}.\footnote{We thank Eldar Ganiev for confirming this observation.} 
It is thus interesting to compare the $q^2$ distributions that follow from the different DM cases.
Because the new particles would add incoherently to the SM rate, we combine the two to obtain a normalized $q^2$ distribution as, 
\begin{eqnarray}
 {d\tilde\Gamma \over d q^2} \equiv 
 { { d \Gamma_{\tt SM} \over d q^2} + { d \Gamma_{\tt NP} \over d q^2} \over \Gamma_{\tt SM} + \Gamma_{\tt NP} }
 = { {d\tilde \Gamma_{\tt SM} \over d q^2 } +(R_{\nu\nu}^K -1 ) {d\tilde \Gamma_{\tt NP} \over d q^2 }  \over R_{\nu\nu}^K },
\end{eqnarray}
where $d\tilde \Gamma_{\tt SM,NP} /d q^2\equiv (d \Gamma_{\tt SM,NP} /d q^2 )/\Gamma_{\tt SM,NP}$.  

\begin{figure}
\begin{center}
\includegraphics[width=5cm]{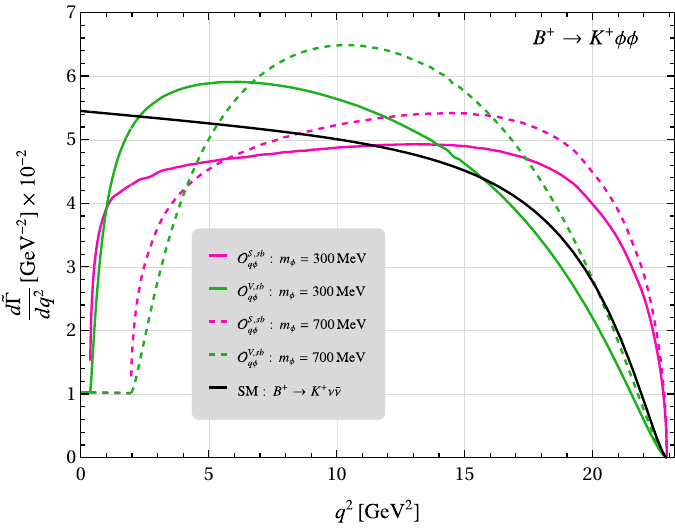} \quad
\includegraphics[width=5cm]{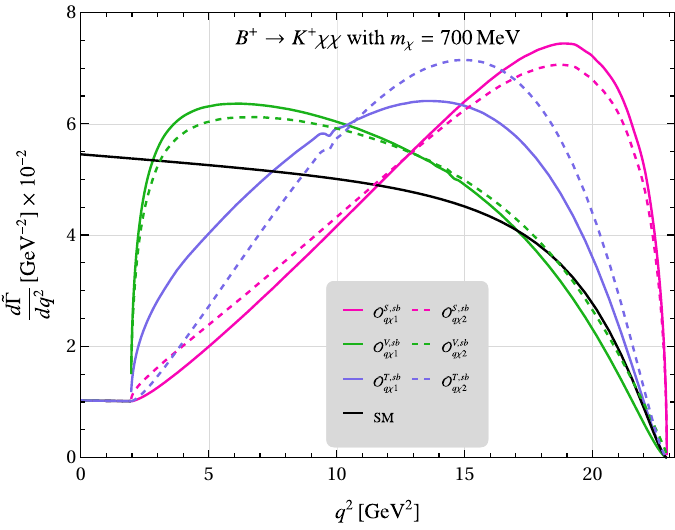} \quad
\includegraphics[width=5cm]{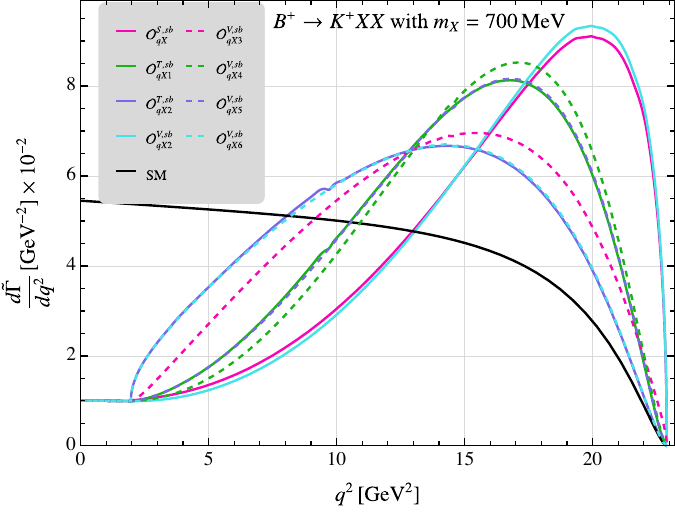} 
\vspace{-0.7cm}
\end{center}
\caption{The $q^2$ distribution of normalized differential decay widths from all three cases: scalar DM (left panel),
fermion DM (middle panel), vector DM (right panel) for selected masses.  }
\label{fig:dGds_dis}
\end{figure}

\begin{figure}
\begin{center}
\includegraphics[width=5cm]{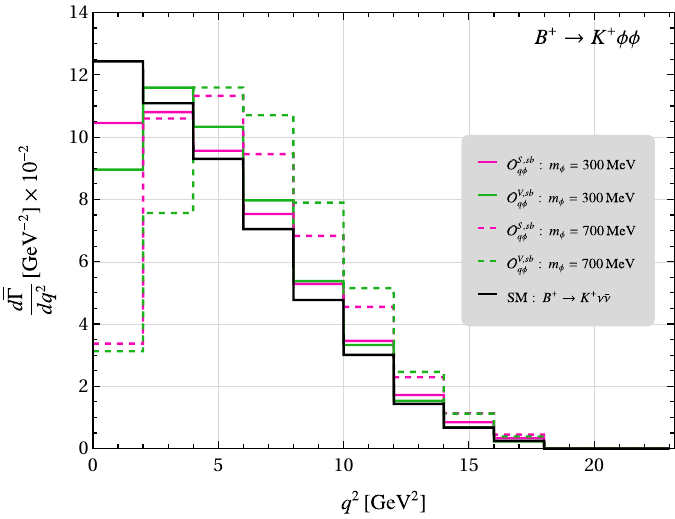} \quad\quad
\includegraphics[width=5cm]{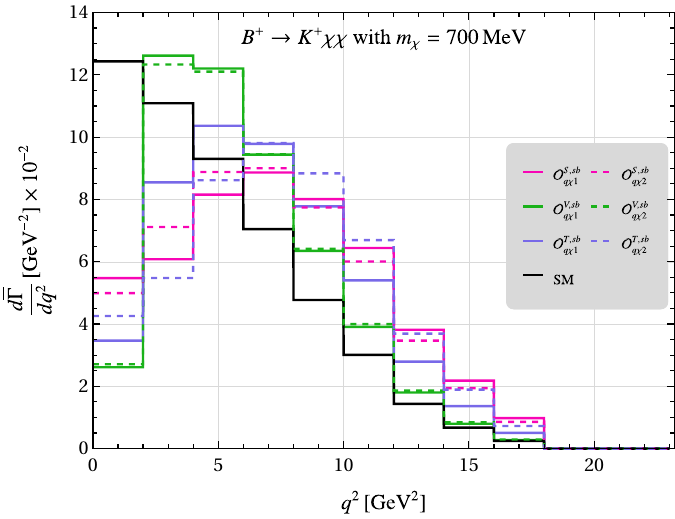} 
\vspace{-0.7cm}
\end{center}
\caption{The $q^2$ distribution of normalized differential decay widths after taking into account the experimental efficiency.
The left panel for scalar DM and right panel for fermion DM.}
\label{fig:dGds_disweff}
\end{figure}

In \cref{fig:dGds_dis}, we show representative cases from the insertion of all the DM operators
discussed above for selected DM mass value $m_{\phi}=300,700\,\rm MeV$ (left panel),
$m_{\chi}=700\,{\rm MeV}$ (centre panel) and  $m_{V}=700\,{\rm MeV}$ (right panel).
For comparison, the SM distribution  is also shown (black line).
The figures indicate that the cases of  $\calO_{q\phi}^{V,sb}$ with $m_{\phi}\sim 300\,\rm MeV$
and $\calO_{q\chi1,2}^{V,sb}$ with $m_{\chi}\sim 700\,{\rm MeV}$ would more closely match the preliminary $q^2$ distribution of the excess. 
This can be seen more clearly when the experimental detection efficiency is taken into account.
\cref{fig:dGds_disweff} shows how the inclusion of the Belle II signal-selection efficiency
for inclusive tagging analysis (ITA) affects the distribution for scalar and fermion DM case.
Since the selection efficiency is given in each $q^2$-bin,
we show the new distributions in histogram according to 
$(d\tilde \Gamma/d q^2)_i\epsilon_i /\left(2\,{\rm GeV^{2}}\sum_i (d\tilde \Gamma/d q^2)_i\epsilon_i \right)$.
Here $\epsilon_i$ is the efficiency taken from Fig.\,6 in \cite{Belle-II:2023esi},
$2\,\rm GeV^2$ is the bin width, and $(d\tilde \Gamma/d q^2)_i$ is evaluated at the central $q^2$ value of each bin.
The study of the shape of this distribution will help narrow down possible explanations if the excess is confirmed.

\section{Summary and Conclusions}
\label{sec:conclusion}

The recent measurement of the branching ratio ${\cal B}(B^+ \to K^+ \nu \bar \nu)$
by the Belle II collaboration is larger than the SM prediction at the $2.7\,\sigma$ level and has attracted some attention.
This new Belle II result is consistent within errors with the average of measurements,
and the discrepancy with the SM is not sufficiently large to require a new physics explanation.
In anticipation of future, more precise, measurements,
the new number invites speculation on the possibility of accommodating a rate that exceeds the SM.
In particular it is important to understand whether models that can enhance $R^{\nu\nu}_K$
can remain consistent with a lower $R^{\nu\nu}_{K^*}$, closer to the SM.

Since the neutrinos in the final state are not identified,
it is possible to enhance $R^{\nu\nu}_K$ with models that contain other invisible particles, such as light dark matter.
It is also possible to do so with additional neutrinos or by introducing lepton flavour violating neutrino couplings
as in models previously discussed in the literature.

We began with a model-independent description of the interactions in $b\to s \nu\bar \nu$.
We considered four types of dimension six operators,
two of which can interfere with the SM (when diagonal in neutrino flavour) and two that can not.
We then scanned the parameter spaces to see if it was possible to cover the new Belle II $1\sigma$ range
for ${\cal B}(B^+  \to K^+ \nu \bar \nu)$ (or the new average) as well as existing bounds on ${\cal B}(B \to K^* \nu \bar \nu)$.
Our results, shown in \cref{f:scan} and \cref{f:lqraa} indicate that whereas
it is relatively easy to reproduce the new $1\,\sigma$ range of $R^{\nu\nu}_K$,
it is much harder to simultaneously obtain a low value of $R^{\nu\nu}_{K^*} <1.9$.
This conclusion is stronger for operators which have no interference with the SM.
The relative size of $R^{\nu\nu}_K$ and $R^{\nu\nu}_{K^*}$ will play an important role in untangling any hint for NP in these modes.

We then restricted the parameter scans to models with one leptoquark,
and found that only $\tilde{S}_{1/2}$ or $V_{1/2}$ can produce $R^{\nu\nu}_K\neq R^{\nu\nu}_{K^*}$
and satisfy the constraints from both $R^{\nu\nu}_{K^*}$ and $b\to s\ell^+\ell^-$ modes.
However, they cannot explain the  $b\to s\ell^+\ell^-$ anomalies.
These two cases predict significant enhancements for the modes  $B^+\to K^+\mu^-\tau^+$,
$B_s\to \mu^-\tau^+$, $B^+\to K^+\tau^-\tau^+$ and $B_s\to \tau^-\tau^+$.
In contrast, we found that models with only $C_L^{ij}$ coefficients,
such as those with $S_0,S_1$ or $V_1$ leptoquarks, cannot satisfy both conditions on $R^{\nu\nu}_K$ and $R^{\nu\nu}_{K^*}$
because they give $R^{\nu\nu}_K= R^{\nu\nu}_{K^*}$.
Furthermore, if $S_0$ is used to explain the current values of $R_{D^{(*)}}$,
it produces values of $R^{\nu\nu}_K$ and $R^{\nu\nu}_{K^*}$ much larger than the existing constraints. 
We also found that solutions with light right-handed neutrinos that
couple to SM fields via a $Z^\prime$ are excluded when $B_s$ mixing constraints are imposed.

In scenarios with light dark matter resulting in $B^+\to K^+ + \slashed{E}$,
we present constraints on the effective scale of all lowest dimensional operators coupling $b,s$ quark bilinears
to pairs of scalar, fermion or vector dark matter particles.
We find that existing experimental upper bounds on $B^0\to K^0 + \slashed{E}$, $B^0\to K^{0*} + \slashed{E}$
and $B^+\to K^{+*} + \slashed{E}$ rule out much of the parameter space
where these operators could enhance $B^+\to K^+ + \slashed{E}$ to the level of the new Belle  II result. 

We illustrated the remaining parameter space that can survive the constraints,
finding that the effective scales are typically constrained to be in the multi-TeV range for scalar and fermion dark matter.
For vector dark matter, the effective scales are in the several hundred GeV range when
the masses are less than 100 MeV and become weaker as the mass increases due to the reduced phase space. 

We confronted the different operators with the preliminary $q^2$ distribution reported by Belle II.
We found that three operators with specific values for dark matter mass would best accommodate the spectrum,
$\calO_{q\phi}^{V,sb}$ with $m_{\phi}\sim 300\,\rm MeV$ and $\calO_{q\chi1,2}^{V,sb}$ with $m_{\chi}\sim 700\,{\rm MeV}$.
Clearly, the experimental result is preliminary, but our exercise illustrates
how the shape of the spectrum can be used to narrow the possible explanations for an excess in the rate.

\acknowledgements

GV and XGH  were supported in part by the Australian Government through the Australian Research Council Discovery Project DP200101470.
XGH was supported by the Fundamental Research Funds for the Central Universities,
by the National Natural Science Foundation of the People’s Republic of China (No. 12090064, No. 11735010, and No. 11985149),
and by MOST 109– 2112-M-002–017-MY3.
XDM was supported in part by the Guangdong Major Project of Basic and Applied Basic Research No. 2020B0301030008
and by the Grant No.~NSFC-12305110.

\appendix

\section{The differential width for $B \to K^{(*)}\chi\chi$ }
\label{sec:dis_fer_DM}

For the light invisible fermionic DM case, based on the parametrization for the hadronic matrix elements
given in \cite{Ball:2004ye,Bharucha:2015bzk,Gubernari:2018wyi,He:2022ljo}, 
the differential decay widths for $B\to K^+(K^0)\chi\chi$ and $B \to K^{*+}(K^{*0})\chi\chi$ transitions
induced from the interactions in \cref{eq:fer_DM_ope} are calculated to take the following general form, 
\begin{eqnarray}
{d\Gamma_{B\to P\chi\chi} \over d q^2} 
& = & 
{\lambda^{1\over2}(m_B^2, m_P^2, s)  \kappa^{1\over2}(m^2,s) \over 384 \pi^3 m_B^3}
\left\{
{ 3(m_B^2 - m_P^2)^2 \over (m_b- m_s)^2 } f_0^2
\left[  (s - 4 m^2)   \left|C_{q\chi1}^{S,sb}\right|^2
+ s  \left|C_{q\chi2}^{S,sb}\right|^2
\right]
\right.
\nonumber
\\ 
& + &
{2  (s + 2 m^2)  \lambda(m_B^2, m_P^2, s)\over s} f_+^2 \left|C_{q\chi1}^{V,sb}\right|^2
\nonumber
\\ 
& + &
{2  \over s} \left[ 6 m^2 (m_B^2 - m_P^2)^2 f_0^2 + (s -4 m^2) \lambda(m_B^2, m_P^2, s)  f_+^2 \right]  \left|C_{q\chi2}^{V,sb}\right|^2
\nonumber
\\ 
& + & 
 { 4  \lambda(m_B^2, m_P^2, s)\over (m_B+ m_P)^2 }  f_T^2
\left[  (s + 8 m^2)   \left|C_{q\chi1}^{T,sb}\right|^2
+ (s - 4 m^2)   \left|C_{q\chi2}^{T,sb}\right|^2
\right]
\nonumber
\\
& - & 
\left.
{12 m (m_B^2 - m_P^2)^2  \over m_b - m_s }  f_0^2  \Im \left[ C_{q\chi 2 }^{S,sb} C_{q\chi 2 }^{V,sb*} \right]
+ {24  m   \lambda(m_B^2, m_P^2, s) \over m_B + m_P}  f_+ f_T
 \Re\left[ C_{q\chi 1 }^{V,sb} C_{q\chi 1 }^{T,sb*} \right]
 \right\},
\\
{d\Gamma_{B\to V\chi\chi} \over d q^2} 
& = & 
{\lambda^{3\over2}(m_B^2, m_P^2, s)  \kappa^{1\over2}(m^2,s)  \over 96 \pi^3 m_B^3 (m_B+ m_V)^2 } V_0^2  
\left[
(s + 2 m^2) \left|C_{q\chi 1 }^{V,sb}\right|^2
+
(s - 4 m^2)  \left|C_{q\chi 2 }^{V,sb}\right|^2 
\right] 
\nonumber
\\
& + &
{ \lambda^{1\over2}(m_B^2, m_V^2, s)  \kappa^{1\over2}(m^2,s)   \over 48 \pi^3 m_B^3 s } 
 \Big\{
(s+8 m^2) \lambda(m_B^2, m_V^2, s) T_1^2 
 \nonumber
 \\
 & + &
\left.
 (s - 4 m^4) \left[ 
 (m_B^2 - m_V^2)^2  T_2^2 
+ { 8 m_B^2 m_V^2 s \over (m_B+m_V)^2 }T_{23}^2  \right]
\right\}
\left|C_{q\chi 1 }^{T,sb}\right|^2
\nonumber
\\
& + &
{ \lambda^{1\over2}(m_B^2, m_V^2, s)  \kappa^{1\over2}(m^2,s)   \over 48 \pi^3 m_B^3  s } 
 \Big\{
  (s- 4  m^2)  \lambda(m_B^2, m_V^2, s) T_1^2 
 \nonumber
 \\
 & + &
\left.
 (s+ 8 m^2) \left[ 
 (m_B^2 - m_V^2)^2  T_2^2 
+ { 8 m_B^2 m_V^2 s \over (m_B+m_V)^2 }T_{23}^2  \right]
\right\}
\left|C_{q\chi 2 }^{T,sb}\right|^2
 \nonumber
\\
& + & 
{m   \lambda^{3\over2}(m_B^2, m_V^2, s)  \kappa^{1\over2}(m^2,s) 
 \over 8 \pi^3 m_B^3 (m_B + m_V) }  V_0 T_1 
 \Re\left[ C_{q\chi 1 }^{V,sb} C_{q\chi 1 }^{T,sb*} \right],
\end{eqnarray}
where $P=K^+,K^0$ and $V=K^{*+}, K^{*0}$. $C_i^j$ is the corresponding Wilson coefficient of the operator ${\cal O}_i^j$.
The K$\ddot{a}$llen function $\lambda(x,y,z) \equiv x^2 +y^2 +z^2 - 2(xy+yz+zx)$ and $\kappa(m^2,s)=\sqrt{1-4 m^2/s}$,
with $m$ being the DM mass while $s$ the invariant mass of DM pair.
$m_B, m_P, m_V$ are the masses of mesons $B^+, P, V$ and $m_b, m_s$ are masses of $b,s$ quarks, respectively.
$f_0, f_+, V_0, f_T, T_1, T_2, T_{23}$ are hadronic form factors associated with different quark currents,
with definitions that can be found in \cite{Gubernari:2018wyi,He:2022ljo}.

\bibliography{refs.bib}

\end{document}